\def\e{\varepsilon}
\newcommand{\lsim}{\raisebox{-.020in}{$\stackrel{<}{{\scriptstyle
\sim}}$}}
\newcommand{\gsim}{\raisebox{-.020in}{$\stackrel{>}{{\scriptstyle
\sim}}$}}
\def\sqr#1#2{{\vcenter{\vbox{\hrule height.#2pt
\hbox{\vrule width.#2pt height#1pt \kern#1pt
\vrule width.#2pt}
\hrule height.#2pt}}}}

\documentstyle[aps,multicol]{revtex}

\begin{document}

\draft

\title{From Collapse to Freezing in Random Heteropolymers}

\author{Carlos J. Camacho$^1$ and Tone Schanke$^2$\\
{\normalsize \it $^1$Facultad de
F\'\i sica, P. Universidad Cat\'olica de Chile, Casilla 306, Santiago
22, Chile}\\
{\normalsize \it $^2$Institute of Physics, Norwegian University of Science
and Technology,
7034 Trondheim, Norway}\\}

\maketitle
\begin{abstract}
{
We consider a two-letter self-avoiding (square) lattice heteropolymer model of
$N_H$ (out of $N$) attracting sites. 
At zero temperature, permanent links are formed leading to collapse
structures for any fraction $\rho_H=N_H/N$. The
average chain size scales as $R\simeq N^{1/d}F(\rho_H)$
($d$ is space dimension).
As $\rho_H\rightarrow 0$, $F(\rho_H)\sim
\rho_H^\zeta$ with $\zeta={1/d-\nu}=-1/4$ for $d=2$. 
Moreover, for $0<\rho_H<1$, entropy approaches zero as $N\rightarrow\infty$
(being finite for a homopolymer). 
An abrupt decrease in entropy occurs at the phase boundary between
the swollen ($R\sim N^\nu$) and collapsed region.
Scaling arguments predict
different regimes depending on the ensemble of crosslinks. Some
implications to the
protein folding problem are discussed

}
\end{abstract}
\bigskip

\begin{multicols}{2}
\columnseprule 0pt
\narrowtext

\newpage

The role of crosslinks in polymers have relevant applications
for many kind of systems like proteins, DNA and other copolymers.
Recently, it has been shown \cite{CJC} that random crosslinking
of residues imposes stringent constraints in the protein 
folding kinetics.
Assuming that there is one ``correct'' set of
crosslinks resembling the native structure, it is found that the
time needed for fast folding sequences to reach this state 
scales as $N^\lambda$, where $N$ is the
number of monomers (or residues) in the chain and 
$\lambda\simeq 3$ ($\lambda \simeq 4$ at the onset).
The model suggests that 
the size of the critical nucleus is on the order of the system size.
Polymer crosslinking is
also important for structure determination using {\it NMR}
\cite{NMR}. This technique 
determines a limited number of contacts in, say, proteins. 
Hence, one would like to
understand how
crosslinks constrain the possible
conformations.
Finally, we mention the process of vulcanization 
where concentrated solutions of crosslinked polymers become
amorphous. These materials 
undergo
a thermodynamic phase transition to a frozen phase if the number
of crosslinks exceeds some critical value \cite{CGZ}.

For these reasons it is desirable to understand the role of
internal constraints in polymers. 
Based on mean-field or {\it ideal} (random walk) polymer
models, recent attempts to address this problem have given
conflicting suggestions. Gutin and Shakhnovich \cite{AGES} found
that the conformational entropy smoothly decreases as 
the number of crosslinks increases. These authors have hinted that 
these conclusions may depend on the ensemble of links.
Bryngelson and Thirumalai \cite{BT}
found a threshold density of links, scaling as $1/\ln N$, beyond which polymers
collapse. On the other hand, it has been found \cite{KK}
that the typical size $R$ of a random walk in $d=2$ dimensions is
reduced by $M$ links to $R\approx (N/M)^{\nu_0}$, where $\nu_0=1/2$ is
the random walk exponent. Kantor and Kardar [6a] conjectured that
for a self-avoiding polymer $R\,\gsim\, (N/M)^\nu$, where $\nu$ is
the standard correlation length
exponent for self-avoiding walks (SAW). 
Accordingly, polymers collapse 
if the number of crosslinks scales as 
$M\sim N^\phi$, with $\phi\,\gsim\, 1-1/d\nu$.

In this letter, we move beyond mean field to show
that when links form freely among
a random set of sites (annealed case) then 
polymers do not collapse to a
compact state, unless the number of constraints
scales linearly in $N$. We should point out that in all likelihood
the annealed case is a better model for real polymers.
To reach this conclusion we analyze the whole sequence space of a
two-letter 
heteropolymer model with $N_H$ ``hydrophobic'' attracting sites
and $N_P$ ``hydrophilic'' (or ``polar'') sites.
The polymer chain
is represented by a SAW
of $N=N_H+N_P$ sites on the square lattice with spacing
$a$. If two H sites 
are nearest neighbors, a short range
attractive energy is assumed adding $-\e<0$ to the
conformational energy of the chain. The only interaction, 
besides the aforementioned attraction between H sites, is 
self-avoidance which forbids two sites from occupying the same site.
It should be mentioned that this model has been extensively used to study
protein folding \cite{CDCOM,PRL,DILL}, there
a limited number of hydrophobic sites are
believed to play a dominant role in the folding process. 
At zero temperature, $H$ sites form permanent links.
Using exact series enumeration \cite{SERIES} 
of all possible crosslinked conformations,
we obtain {\it exact} thermodynamic quantities for $N \le 20$.
Analytically, we consider Flory's affine network theory of
rubber elasticity to generalize some of our conclusions to
the problem of quenched random links. 
In what follows, we work in adimensional units with
$a$, $\e=1$.

It is well known 
that at some critical ``theta'' temperature $T_\theta$ homopolymers
undergo a coil-to-globular (or collapse) transition.
Below $T_\theta$ polymers
collapse to an
average radius of gyration $\langle R_G\rangle$ scaling
as $N^{1/d}$.
Above $T_\theta$ polymers
are swollen (or extended) with $\langle R_G\rangle\sim N^\nu$, 
where $\nu=3/4$ and $0.59_2$ for $d=2$ and 3 \cite{TG}, respectively.
To analyze heteropolymers 
we proceed by computing the radius of gyration $\langle \overline{R_G^2}
\rangle$, where the upper bar means average over sequence space, i.e.
$N\choose N_H$ sequences for any given $N$ and $N_H$.
Although the data shown in this letter corresponds to $d=2$, it is
helpful theoretically to keep the symbols $d$ and $\nu$ in evidence.

As shown in Fig. 1, we find that, at $T=0$, $\langle \overline{R_G^2}\rangle$
is very well described
by the scaling law
\begin{equation}
{\langle \overline{R_G^2}\rangle\over N^{2/d}} \approx F(\rho_H),
\hskip.4cm\hbox{with}\hskip.4cm \rho_H={N_H\over N}\,,
\end{equation}
where $\rho_H \equiv 1-\rho_P$ corresponds to the fraction of H sites. 
We note
that in general one should have allowed the scaling variable $\rho$
to depend on a suitable crossover exponent, say, $\rho_H =N_H/N^\phi$. 
Our results,
however, indicate that $\phi =1$.
Eq. 1 demonstrates that polymers collapse if and only if $N_H$ scales
as $N$. 
This result is clearly not obvious. Namely,
hydrophilic chains with a tiny, but {\it finite}, fraction of randomly  
distributed attracting sites are collapsed at $T=0$.

Furthermore, the scaling function $F(\rho_H)$ is expected to have
well defined asymptotic laws in both the {\it hydrophilic} 
$\rho_H\rightarrow 0$, and the {\it hydrophobic} limit $\rho_P\rightarrow 0$.
For $\rho_H\rightarrow 0$, one should recover the self-avoiding walk
exponent (as a function of $N$). Hence,
\begin{equation}
F(\rho_H)\approx A\rho_H^{2\zeta}, \hskip.4cm\hbox{with}\hskip.4cm 
\zeta={1/d-\nu}.
\end{equation}
This is in excellent agreement with the slope $\zeta=-1/4$ observed
in Fig. 1.
At $T=0$, the chain ensemble corresponds to that
of maximally crosslinked chains. Indeed, we find 
$\langle \overline{M}\rangle\sim N_H$ \cite{UNP}, suggesting
the validity of (1) with a scaling variable $\rho=M/N$ 
for $M$ annealed random links.
Based on (2), we predict $\zeta\simeq -0.25_9$ for $d=3$.
It is noteworthy that if one fixes 
$N_H$ and $N\rightarrow\infty$, then the transition
between SAW behavior and the collapse
regime occurs at 
$\rho_H^*=1-\rho_P^*\simeq 0.61$ (see Fig. 1). 

The hydrophobic limit is shown in the inset of Fig. 1.
For $\rho_P\rightarrow 0$, chains collapse, 
approaching a sphere of volume 
$V\approx Na^d + O(N^{\sigma/d})$, 
where $a^d$ is volume of lattice cell.
Naively, we might expect $\sigma$ to be a surface correction,
{\it i.e.} $\sigma=d-1$. The data, however, shows
\begin{equation}
{\langle \overline{R_G^2}\rangle\over N^{2/d}} \approx R^2_0 + B\rho_P^
{(d-\sigma)/d}\,,
\end{equation}
where $R^2_0\equiv a^2/(2\pi)$, 
and $\sigma=0.7\pm 0.1$ ($d=2$) is a {\it novel} universal exponent
describing the approach to circularity of a collapsing chain.
Interestingly, (3) is related to the longstanding
problem of how many lattice points
fit inside a sphere of volume $V$, where $\sigma$ is known to
vary between 1/2 and (upper bound) $7/11<1$ \cite{SECO}.
The slope in Fig. 1 (inset) corresponds to the scaled
version of this exponent.
For $d=3$, 
$R^2_0\equiv (3a^3/4\pi)^{2/3}3/5$ and $\sigma < 2$ \cite{SECO}.
The apparent deviations from scaling 
at $N_P\rightarrow 0$ are well known finite-size effects on
the shape of collapsed lattice chains
\cite{CDCOM,PRL}. 

A similar analysis of the conformational entropy $\bar s_0(\rho_H)\equiv
\overline{\ln\Omega(\rho_H)}/N$, where $\Omega$ is number
of conformations, leads to the scaling plot shown in Fig. 2.
In the hydrophilic region $\rho_H\,\lsim\,0.6$, 
$
\bar s_0(\rho_H)\approx G(\rho_H)/N^\chi$, 
with $\chi=0.43\pm0.04$.
For $\rho_H\,\gsim\,0.6$, entropy decreases even faster with $N$.
In the SAW limit $\rho_H\rightarrow 0$, $\bar s_0$ approaches
a constant yielding $G(\rho_H)\sim\rho_H^{-\chi}$. 
Scaling breaks down due to finite-size effects at 
$\rho_P\sim O(N^{-1/d})$. At this point, hydrophilic sites
rearrange on the surface of the structure and entropy approaches
a constant 
---$\bar s_0(\rho_H=1)=\ln{q/e}$, 
where $q$ is the coordination number of the lattice \cite{PRL,GOT}.
Hence, in sharp contrast with the homopolymer cases $\rho_H=0$ and $\rho_P=0$, 
{\it chains have zero entropy and are collapsed} for
any finite $\rho_H<1$ and $N\rightarrow\infty$. Indeed, the
{\it internal} network of permanent links 
formed for $N_H\sim O(N)>N^{(d-1)/d}$
yields enough constraints to change the qualitative properties 
of polymers.
We should mention that the entropy of maximally  compact 
structures in heteropolymers
have already been shown \cite{PRL}
to have a deep minimum around $\rho_H\simeq 0.6$.

As a function of temperature,
from the homopolymer (theta) case, we start increasing
$\rho_P$ reducing the overall drive towards collapse.
Then, as indicated in the phase diagram of Fig. 3, 
the collapse transition temperature $T_x(\rho_P)$ ---which
divides the swollen from the collapsed region--- goes down, and
eventually to zero at $\rho_P=1$.
Kantor and Kardar \cite{KK1} have shown a related phase diagram for
random ``charges'' in a $d=3$ chain. On a regime
where {\it both} $H$ and $P$ sites attract each other, they found
that chains collapse for $|N_H-N_P|/N$ between 0 and 1. Note, however,
that their model cannot sample the hydrophilic regime 
with less than 50\% of attracting sites.

Entropy is $\bar s(\rho_H,\,T)=(\overline{E}-
\overline{F})/TN$, where
$E$ and $F$ are the energy and free energy, respectively.
Upon crossing $T_x(\rho_P)$,  heteropolymer chains
have a sharper decrease in entropy than homopolymers (see inset
in Fig. 2). This sharpness appears to be higher than what would be 
expected for a critical transition. Moreover, 
the remanent entropy below
$T_x(\rho_P)$ decreases with system size, whereas above $T_x(\rho_P)$
it remains constant. All these suggest
that the nature of the transition changes from
critical to 1$^{st}$ order at some tricritical point $T_x(\rho_P^c)$
---from our limited data, we conjecture $0<\rho_P^c < 0.4$ 
(see Fig. 3). Furthermore, one could also argue that given the first order 
jump in entropy 
at $T=0$ and $\rho_P=1$ ($N\rightarrow\infty$), 
by {\it continuity} it is reasonable
to expect a line of first order transitions beginning at $T=0$ and
going to finite temperatures. 
Similar phase diagrams have been obtained for, say, 
the tricritical point in
a dilute magnet \cite{Grif}. The analogy here is diluting 
a  ``theta'' polymer.
It is also worth mentioning that a closely related model solved by
Garel et. al. \cite{GOT} shows that
the nature of the collapse 
transition depends on the 
hydrophobic-hydrophilic content of the chain. In the hydrophilic regime,
the transition is first-order. Whereas
in the {\it strong} hydrophobic regime, the transition
is continuous similar to an ordinary ``theta'' point.
An impressive, and yet intriguing,
result is that even if the collapse transition is continuous,
the low temperature phase as no entropy (except at
$\rho_P=0,1$). 

In the hydrophilic regime $\rho_H\,\lsim\,\rho_H^*$ ($T=0$),
the average number of conformations $\bar\Omega(\rho_H)$ 
grows exponentially in $N$,
changing to non-exponential growth only at
$\rho_H^*\simeq 0.61\pm 0.05$. The change
in the scaling of $\bar\Omega(\rho_H)$ is rather abrupt. 
Below $\rho_H^*$, $\ln{\bar\Omega(\rho_H)}/N$
increases as $\sim (\rho_H^*-\rho_H)^\omega$,
with $\omega\simeq 0.5\pm 0.1$ \cite{UNP}. 
Above $\rho_H^*$, 
$\ln{\bar\Omega(\rho_H)}/N$
consistently decreases towards zero.
This means that the structural localization in some few structures
is particularly
strong at and below $\rho_H^*$, see also
Fig. 2. It is tempting to speculate that this special
point $\rho_H^*$ is related
to a rigidity percolation transition (see, e.g., \cite{Jac})
from
a rigid to a floppy structure, or
to a ``vulcanization'' transition of a single chain. 
Certainly, these aspects of the model deserve further study.

For completeness, we assess the question: ``What happens for
crosslinks that can form arbitrarily apart along the
backbone (quenched case)?''
A general analysis of crosslinks in polymers
can be made by means of Flory's affine network theory of
rubber elasticity. In this framework, the total free energy
of a polymer of $N$ sites and $M_4$ crosslinks of functionality four
can be constructed (see, {\it e.g.}, \cite{LB}).
By considering an elastic, repulsive and entropic energy term, plus
an ideal gas as solvent, the typical size of a
crosslinked polymer is found to be  \cite{UNP}
\begin{equation}
R\sim \rho^{-1/(d+2)} N^{2/(d+2)}\hskip.4cm\hbox{with}\hskip.4cm
\rho=M_4/N\,.
\end{equation}
This expression is expected to be valid for $\rho\ll 1$. Strikingly,
as $N\rightarrow\infty$, we automatically recover
Flory's exponent for a SAW with $R\sim N^{3/(d+2)}$. Moreover, for $d=2$
the theory predicts
the same scaling form and exponents as in Eqs. 1 and 2,
with $R\sim \rho^{-1/4} N^{1/d}$.

For $d=3$, a new scaling behavior is predicted, namely
$
R\sim \rho^{-1/5} N^{2/5}$.
For a given ratio $\rho$
conformations are neither
fully collapsed nor swollen \cite{CJC}. This behavior has also been implied
by Levin and Barbosa \cite{LB} in a
study of phase transitions of neutral
polyampholyte. The prediction is
that polymers collapse if $M_4\sim N^\phi$, with
$\phi=4/3$. However, if we also allow sites
with functionality larger than four, then we get back to $\phi=1$
as in (1).
Hence, it is much harder to collapse a chain with
two-particle links than with links that do not saturate.

In summary, the theoretical and numerical study of
heteropolymers and random crosslinked chains has
revealed a variety of different regimes, some
with simple scaling behavior, others more complex and
intriguing.
For $d=2$, we find that polymers collapse if and only if
the number of mutually attracting monomers $N_H=N-N_P$ scales as
the size of the system $N$. At zero
temperature, novel universal
exponents describe the limiting behavior for $\rho_H=N_H/N\rightarrow 0$
and 1. We expect the same conclusion to be valid for $d=3$.
For $d=2$, the problem of annealed and quenched random links
are predicted to have the same scaling properties.
For $d=3$, quenched (two-particle) links collapse a chain if $M\sim N^{4/3}$.
The nature of the collapse transition changes
from first to second order for some small enough density of
non-interacting sites $\rho_P=N_P/N$.
We note that almost simultaneously with collapse there is
an abrupt decrease in entropy. Tracing this entropic change
to a folding transition suggests a favorable
scenario to find fast folding proteins \cite{CJCDT}.
In the thermodynamic limit ($T=0$) random HP
chains have zero entropy for any finite fraction $\rho_H<1$.
This entropic crisis and collapse is due
to the network of {\it internal} constraints buried in the
structure.
This result has yet to be fully understood in the context
of random heteropolymers with quenched disorder.
Properties on the size, entropy, and number of conformations as
a function of $\rho_H$ indicate
the existence of a special point at 
$\rho_H^*\simeq 0.61$, most likely related to a rigidity
percolation or vulcanization transition. 
Our results show that both structural determination and collapse require
a relatively large number of constraints ($\gsim\, N$) 
suggesting that
the thermodynamics and dynamics of crosslinking should
play an important role in protein folding \cite{CJC}.

\centerline{***}

We are indebted to F. Claro for the stimulating role he played
in this project. We are grateful with M.E. Fisher and M. Kardar for helpful
correspondence. CJC wish to thank T. Garel for many informative
conversations and valuable suggestions to the manuscript.
This work was supported in part by FONDECYT Nrs. 1930553 and 3940016,
and the Cluster de Computaci\'on Cient\'{\i}fica at PUC (Chile). 
T.S. would like to thank Universidad Cat\'olica for
its hospitality during her internship, when this work was done.

\begin{figure}
\caption{{\it Scaling of squared radius of gyration} averaged
over sequence space
as a function of 
fraction of hydrophobic (attracting) sites $\rho_H=N_H/N$, and 
of (inset) hydrophilic (non-interacting) sites $\rho_P=N_P/N$.
Points with $N_H=0$ and 1
are not shown as they correspond to unrestricted SAW.
Symbols 
correspond to exact data for the square lattice. The limits for
$\rho_H$ and $\rho_P\rightarrow 0$ are indicated by the dashed lines, i.e.
$2\zeta=-1/2$ and 
$A = .132\pm .001$ (2), and (inset) $\sigma=0.7$ and $B=.036_4\pm.003$ (3),
respectively. (For $N \ge 18$, some data points with $\rho_H\sim 0.5$
are missing due to CPU constraints.)}
\end{figure}

\begin{figure}
\caption{ {\it Entropy} $\bar s_0(\rho_H)$
as a function of the scaling
variable $\rho_H$, at $T=0$. Same symbols 
as in Fig. 1. Inset, entropy $\bar s(T)$ as a function of temperature
for $N=15$ and $N_P=0$ (dashed line) and
$N_P>0$ (solid lines). Values of $N_P$ are indicated in figure, 
see also dotted lines in Fig. 3. 
The collapse temperature $T_x(\rho_P)$
is indicated by the square symbols. The curves for $N_P=7$ and
10 are indicative of a sharper transition than the $N_P=0$ case.
}
\end{figure}

\begin{figure}
\caption{Schematic phase diagram of a heteropolymer. Square symbols show
exact position of peak in  the 
energy fluctuations $\Delta \bar E=\langle \bar E^2\rangle-
\langle \bar E\rangle^2$  for $N=15$ and 
$N_P = 0,\cdots,13$. 
Our data suggests the possibility of a first-order collapse transition
(solid line)
for hydrophilic chains changing to second-order (dashed line) for
hydrophobic chains at some point
$T_x(\rho_P^c)$, see text. 
A solid circle indicates the position of $\rho_P^*$.
Curves in Fig. 2 are computed along dotted lines. 
}
\end{figure}

\vfill\break

\end{multicols}

\end{document}